\documentclass[%
reprint,
 amsmath,amssymb,
prl,
]{revtex4-2}
\usepackage{graphicx}
\usepackage{dcolumn}
\usepackage{bm}
\usepackage{gensymb}

\graphicspath{ {./} }
\begin{document}

\preprint{APS/123-QED}

\title{The impact of electron inertia on collisional laser absorption for high energy density plasmas}

\author{James R. Young}
\author{Pierre-Alexandre Gourdain}%
\affiliation{%
 Department of Physics and Astronomy, University of Rochester, Rochester, New York 14627, USA\\
}%
\affiliation{Laboratory for Laser Energetics, University of Rochester, Rochester, New York 14627, USA\\.}

\date{\today}

\begin{abstract}
High-power lasers are at the forefront of science in many domains. While their fields are still far from reaching the Schwinger limit, they have been used in extreme regimes, to successfully accelerate particles at high energies, or to reproduce phenomena observed in astrophysical settings. However, our understanding of laser plasma interactions is limited by numerical simulations, which are very expensive to run as short temporal and spatial scales need to be resolved explicitly.
Under such circumstances, a non-collisional approach to model laser-plasma interactions becomes numerically expensive. Even a collisional approach, modeling the electrons and ions as independent fluids, is slow in practice. In both cases, the limitation comes from a direct computation of electron motion. In this work, we show how the generalized Ohm's law captures collisional absorption phenomena through the macroscopic interactions of laser fields, electron flows, and ion dynamics. This approach replicates several features usually associated with explicit electron motion, such as cut-off density, reflection and absorption. As the electron dynamics is now solved implicitly, the spatial and temporal scales of this model fit well between multi-fluid and standard magnetohydrodynamics scales, allowing to study a new class of problems that would be too expensive to solve numerically with other methods. 

\end{abstract}

\maketitle

\section*{introduction}

Laser-plasma interactions (LPI) are key features when a high-power laser drives materials into ionized states. For example stimulated Brillouin  \cite{kruer_physics_2003} combined with ponderomotively driven plasma response has long been a major concern for inertial confinement fusion (ICF) \cite{bezzerides_convective_1996, eliseev_stimulated_1995}, especially when cross-beam energy transfer (CBET)\cite{huller_crossed_2020, ruocco_modelling_2020} is involved. In addition to providing a source of magnetization in plasmas\cite{gradov_magnetic-field_1983, shukla_magnetization_2009}, the very nature of LPIs can change in the presence of strong magnetic fields\cite{ivanov_generation_2021}, like those generated around magnetars\cite{palmer_giant_2005} and white-dwarves \cite{jordan_fraction_2007}, and prior research even links fast radio bursts in the relativistic winds of magnetars\cite{sobacchi_saturation_2023, ghosh_nonlinear_2022} to ponderomotive self-focusing of electromagnetic radiation\cite{kaw_filamentation_1973, del_pizzo_self-focussing_1979}.  LPI can also cause electron and ion density modulations that grow and steepen\cite{estabrook_two-dimensional_1975, dragila_laser_1978, smith_particle--cell_2019} and whose sharp nature has been recently used to form plasma gratings\cite{lehmann_plasma_2019, sheng_plasma_2003, plaja_analytical_1997} for high-power plasma optics.  These can be produced through ionization effects of overlapping laser beams \cite{edwards_plasma_2022,edwards_control_2023, shi_generation_2011} or ponderomotively driven ion and electron peaks \cite{sheng_plasma_2003, peng_dynamical_2020, peng_nonlinear_2019}.  The former requires models capable of handling not only ionization, but also the effect of the laser on the newly formed under-dense plasma, while the latter requires resolving electron effects on the ion scale.  For this work we focus on collisional absorption, which is the primary ingredient of inverse Bremsstrahlung (IB)\cite{kruer_physics_2003, dawson_highfrequency_1962} and is the most fundamental laser-plasma interaction.   

 In one regime, when the ions are mostly frozen in place, a common situation in laser wakefield acceleration simulations, particle-in-cell models \cite{hockney1981} capture LPIs extremely well. In the other regime, when the ion and electron motion is ambipolar, the essence of LPIs is deeply rooted in IB, where ray-tracing (RT) \cite{kaiser_laser_2000, colaitis_inverse_2021} can be used to track the energy budget left inside the electric field as the laser beam travels inside the plasma.  This model rests on the application of WKB approximation, which assumes the electron density varies slowly enough to have a well-defined wave-vector and frequency.  While the first approach is typically collisionless (although collisional models have been successfully added \cite{sentoku_numerical_2008}) and uses the Vlasov approximation, the second is collisional and uses conservation laws to track the ion dynamics\cite{braginskii1965}. State-of-the-art MHD codes (e.g. \textit{FLASH} \cite{fryxell_flash:_2000}, \textit{LILAC} \cite{delettrez_effect_1987}, \textit{DRACO} \cite{keller_draco_1999_apsb}, \textit{HYDRA} \cite{marinak_threedimensional_1996}) often drop some or all of these electron terms. They split the LPI into a radiation component (eikonal approximation of the laser field) and an absorption component (i.e. IB) while only coupling the electron and rays via the plasma dispersion relation \cite{stix1992waves}. Unfortunately, problems that fall in between regimes are more difficult to model. Hybrid codes \cite{HOLDERIED} have been used recently to bridge this scale gap, running a multi-fluid approximation when possible and switching to kinetic when the fluid approach breaks down.

However when LPIs occur in a fully formed plasma \cite{young_using_2021}, such as those produced in modern pulsed-power machines \cite{shapovalov_oil-free_2017, shapovalov_low-inductance_2019, zucchini2015,strucka2022}, the ray approximation usually breaks down. The large density gradients and complex geometrical structures present before the laser is fired will generate situations where the paraxial assumption is violated and ray-tracing cannot be used. Additionally, ray-tracing does not inherently include interactions between rays, which is necessary to model more complex LPI such as CBET.  Additional models must be used to allow ray interactions \cite{follett_ray-based_2018, harvey_modeling_2015}.  Resonant absorption for p-polarized light is not natively captured by ray tracing and requires analytic asymptotic solutions for the absorption coefficients\cite{hinkel-lipsker_analytic_1989}.  For hot electron temperatures and large spatial density gradients, this can reduce the collisional (or Landau) damping by as much as 50\%\cite{michel_introduction_2023}, which makes this an important subject for ICF\cite{freidberg_resonant_1972, palastro_resonance_2018}.

Up to now, such situations relied on blending multiple computational approaches to study these regimes\cite{davies_laser-driven_2017, hansen_measuring_2018}. This letter shows that the physics necessary to capture collisional LPI is fully accounted for in XMHD, with the generalized Ohm's law (GOL), where electron terms have been retained. Since correctly capturing IB only relies on collisional LPI, then at least for low energy lasers, the primary energy delivery mechanism is natively present.  We expect this to transition away from IB when the laser power is increased, which will require additional forms of LPI. While we restrict our discussion to non-relativistic plasma interaction using high power (i.e. long pulse) lasers, this framework can be extended to relativistic plasmas by modifying the conservation laws and the GOL \cite{hamlin_relativistic_2016}. 

The first section uses dimensional analysis to show which terms in the GOL are important to LPIs in the extended magnetohydrodynamics (XMHD) framework. Then, we show quantitatively how XMHD handles LPIs in slab geometries using PERSEUS \cite{seyler_relaxation_2011, martin_2010}. Finally, we highlight how energy transfers from the laser field to the plasma. 

\section*{The Generalized Ohm's law scaled for laser-plasma interactions}

\subsection*{Redimensionalization of the generalized Ohm's law using LPI characteristic scales}
Dimensional analysis comes into play to make important parameters appear in XMHD equations \cite{seyler_relaxation_2011}. Removing all units in conservation laws and Maxwell's equation can be done with taking three independent characteristics scales such as time $t_0$, speed $v_0$ and mass density $\rho_0$ \cite{gourdain_impact_2017}. One key step is necessary to arrive to a self consistent result. Conservation of momentum dictates that the characteristic speed $v_0$ should also be the characteristic Alfv\'en speed. In the framework of laser plasma interactions, we can take the electromagnetic wave speed $c$ as our characteristic speed, $v_0=c$. Following Faraday's law, we relate the electric and magnetic fields characteristic scales as $E_0=cB_0$. As we remove dimensions from each XMHD equation, we find the following relationships between the different characteristic scales:
the thermal pressure scale is $p_0=\rho_0c^2$, the resistivity scale $\eta_0=t_0/\varepsilon_0$,
the magnetic field scale is $B_0=\sqrt{\rho_0/\varepsilon_0}$, and the current density scale is $j_0=\sqrt{\rho_0/\mu_0}/t_0$.

We assume here quasi-neutrality and separate electron and ion energies. However the electron momentum is computed implicitly, using the time evolution of the electrical current density via the GOL, given by

\begin{eqnarray}\label{eq:dimensionless_gol}
    \mathbf{E}=-\mathbf{v} \times \mathbf{B}+\eta\textbf{j}+\frac{1}{\omega_{pe}t_0}\sqrt{\frac{m}{n_e}}(\textbf{j}\times\textbf{B}-\vec\nabla p_e)\nonumber\\ 
    + \left(\frac{1}{\omega_{pe}t_0}\right)^2\left(\frac{\partial \mathbf{j}}{\partial t}+\vec\nabla \cdot (\mathbf{vj+jv})\right)\\
    -\left(\frac{1}{\omega_{pe}t_0}\right)^3\sqrt{mn_e}\,\vec\nabla\cdot\left(\frac{1}{n_e}\mathbf{jj}\right).\nonumber 
\end{eqnarray}
Here \textbf{E} is the electric field, \textbf{v} is the flow speed, \textbf{B} is the magnetic field, \textbf{j} is the current density, $n_e$ is the electron number density, $m$ is the ratio of the ion mass to the electron mass, $\rho$ is the mass density, $p$ is the thermal pressure, and $\eta$ is the electric resistivity, \textit{all dimensionless}.

The dimensional electron plasma frequency $\omega_{pe}=\sqrt{N_ee^2/(m_e\varepsilon_0)}$ is highlighted by the nondimensionalization process. Further, $N_e$ is electron number density, e is the elementary charge and $m_e$ is the electron mass,  \textit{all dimensional}. It is now in front of all the terms connected to electron physics inside Eq. \eqref{eq:dimensionless_gol}.

\subsection*{Characterization of LPIs in XMHD}

To understand what type of LPI takes place in XMHD, we can look at the interaction of a plane electromagnetic wave traveling to the left with a plasma slab located in the region $x<0$. At early times, the ion fluid is static and has no spatial dependence in the region $x<0$. We can equate the laser electric field at the plasma-vacuum interface, the LHS of Eq. \eqref{eq:dimensionless_gol}, to the electric field inside the plasma, the RHS of Eq. \eqref{eq:dimensionless_gol}. In this case, Eq. \eqref{eq:dimensionless_gol} simplifies to
\begin{equation}
    \mathbf{E}\approx\eta\textbf{j}+ \frac{1}{\omega_{pe}^2t_0^2}\frac{\partial \mathbf{j}}{\partial t},
\label{eq:simplified_dimensionless_gol}
\end{equation}
dropping the third order terms in $1/\omega_{pe}t_0$ since the characteristic time scale of the interaction $t_0$ is much larger than the electron time scale (i.e. $1/\omega_{pe}$). In other terms, we look at effects that are much slower than the electron plasma frequency. At this early point in time the ions have not moved yet and the Lorentz force equilibrates exactly the electron pressure gradient. So we also drop the first order term in $1/\omega_{pe}t_0$, since 
\begin{equation*}
    \mathbf{j}\times\mathbf{B}=\vec \nabla p_e,
\end{equation*}
\noindent from the momentum equation.  

Modifying Maxwell-Amp\`ere's law using Eq. (\ref{eq:simplified_dimensionless_gol}) while taking $\eta$ and $\omega_{pe}$ to have small time variations compared to $t_0$, we get

\begin{equation}
    \frac{1}{\omega_{pe}^2t_0^2}\frac{\partial^2 \mathbf{j}}{\partial t^2}+\eta\frac{\partial\textbf{j}}{\partial t}+\textbf{j}\approx\nabla\times\textbf{B}
\label{eq:resonance_dimensionless}
\end{equation}

Eq. (\ref{eq:resonance_dimensionless}) is the equation of a damped harmonic oscillator with a resonant frequency,  $\omega_{pe}$, driven by the curl of the laser magnetic field \textbf{B}. 

 The dimensional analysis shows that LPIs are mediated by electron inertia and Maxwell-Amp\`ere's law in XMHD.  So, the laser energy is transferred to the electron fluid via harmonic oscillation. This is possible because we can now store mechanical energy inside the electron fluid, a transformation that requires the electrons having mass (hence inertia). Resonance allows this energy to become large enough so it can interact non-linearly with the incoming electromagnetic wave. Consequently, XMHD recovers some common features found in the two-fluid models\cite{braginskii1965,chapman1990}, while resolving the electron motion on the ion time scale only.  We refer the reader to prior work \cite{seyler_relaxation_2011, young_using_2021} to see a complete set of equations being solved in the code. 

Although electron inertia has been important in studies of collisionless magnetic reconnection \cite{birn_reconnection_2007, drake_breakup_1997, jones_including_2003, munoz_new_2018}, often the components are broken to include only the convective acceleration term ($\mathbf{v_e} \cdot \nabla \mathbf{j}$) rather than considering temporal variations of current directly \cite{kuznetsova_kinetic_1998}, which are key to absorption and eventual collisional dissipation.

\section*{Validation of XMHD simulations using PIC and RT}

\begin{figure}[!ht]
\centerline{\includegraphics[]{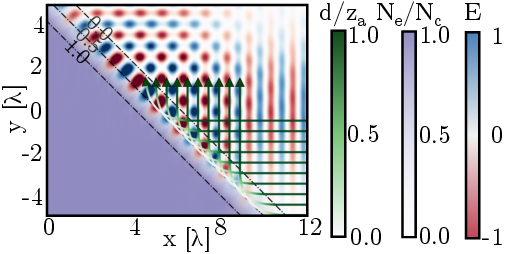}}
\caption{\label{fig:rt_perseus_comp}  A plot of a $\mathrm{45\degree}$ ramp electron density profile with S-polarized light coming from $\mathrm{x=12\lambda}$ with $\mathrm{I_{laser}=3.5\times10^{17}\frac{W}{m^2}}$ and $\mathrm{\lambda=527nm}$.  Overlayed on this are 10 ray-traced solutions for this same density profile that are colored based on their distances from the turning point in units of Airy skin depth, $\mathrm{z_a}$.} 
\end{figure}

\begin{figure}[!ht]
\centerline{\includegraphics[]{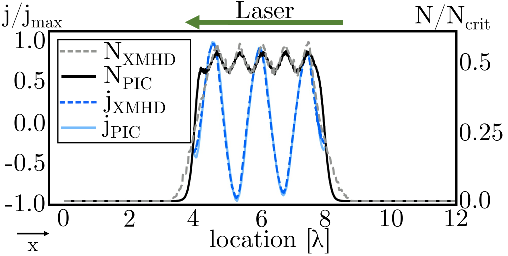}}
\caption{\label{fig:pic_perseus_comp}  A comparison at $t=5.8$ps of current density from XMHD and OSIRIS \cite{hemker_particle_cell_2015} simulations for a 1D plasma initialized with $\mathrm{T_e=1eV}$, $\mathrm{n_e/n_{crit}=0.5}$, $\mathrm{I_{laser}=3.5\times10^{17}\frac{W}{m^2}}$, and $\mathrm{\lambda=527nm}$.  The PIC domain used 8192 particles-per-cell and both simulations used heavily over-sampled cells with $\mathrm{\Delta x=\lambda/120}$.  Also of note are the ponderomotively driven density modulations.} 
\end{figure}

First, we demonstrate this model is consistent with ray-tracing.  Further, XMHD simulations show almost perfect agreement with OSIRIS for a duration of at least 3 orders of magnitude larger than the laser period.  For the XMHD simulations, a non-relativistic ($\mathrm{a_o < 0.01}$) s-polarized laser was propagated into a simulation domain by setting an oscillatory electromagnetic field as a boundary condition.  The magnetic field component, $B_{y}$, was defined on the vertical boundary so the Poynting flux was directed to the left.  The beam has a uniform spatial profile, $\mathrm{B_{y}=B_{max}\cos{\left(kx+\omega t\right)}}$.  Here, $\mathrm{E_{max}=\sqrt{2I_{laser}/(c\epsilon_0)}}$, $\mathrm{k=2\pi/\lambda}$, and $\mathrm{\omega=ck}$ with $\mathrm{I_{laser}}$ varying from $\mathrm{10^{16}\rightarrow 10^{18} \text{Wm}^{-2}}$ ($a_0\approx0.001\rightarrow0.01$) and $\mathrm{\lambda=527 \text{nm}}$.  The PIC simulation uses identical geometry and laser characteristics.  For ray-tracing, we specifically use a ramp density profile from $n_e/n_{crit}=0\rightarrow0.5$.  All simulations other than the RT comparison in Fig. \ref{fig:rt_perseus_comp} are performed in a 2D pencil domain (9 cells in the y-direction and the rest of the resolution in the x-direction).  To see a fully 2D simulation of LPI then we refer the reader to prior work \cite{young_using_2021}.

Fig. \ref{fig:rt_perseus_comp} shows rays as green arrows overlaying the XMHD simulation for an s-polarized laser propagating towards the density ramp.  The WKB approximation assumes the turning point should occur when $\mathrm{n_e/n_{crit}=cos^2(\theta)}$, which for this geometry ($\mathrm{\theta=45\degree}$) is $\mathrm{n_e/n_{crit}=0.5}$.  The color of the green ray shows the distance of the ray from the theoretical turning point in units of Airy skin depth ($\mathrm{z_a}$), within which, the WKB approximation is no longer valid\cite{michel_introduction_2023}.  We see that within one $\mathrm{z_a}$, the light becomes evanescent and decays to zero. For s-polarized light, there is no resonant absorption and the total deposited energy is equivalent to ray-tracing approximation.  However, this will no longer be true for p-polarized light.  

Now we show XMHD also matches PIC simulations.  Fig. \ref{fig:pic_perseus_comp} shows $\mathrm{n_e/n_{crit}}$ and current density, $\mathrm{j}$, for equivalent initializations and boundary conditions for both PIC and XMHD simulations.  XMHD resistivity, $\eta$, was set artificially low to approach a collision-less condition like PIC models.  The output is taken approximately 5.8 picoseconds into the simulation, where we see that not only does $\mathrm{j}$ look identical for both models, but the sinusoidal $\mathrm{n_e/n_{crit}}$ modulation is also remarkably close.  The density oscillations are direct result from ponderomotive effects on the ions, which will be discussed in a future paper.  

We have so far demonstrated compatibility with ray-tracing models as well as a remarkable similarity in late-time simulations with PIC.  So we now present and discuss the mechanisms that lead to collisional energy transfer in XMHD.  

\section*{Collisional absorption in LPI}
This work will now focus on an under-dense regime as this allows us to investigate both the resonance in Eq. \ref{eq:resonance_dimensionless} and collisional absorption within a single domain.  Since XMHD agrees with PIC for long time durations, we focus here on shorter times to catch the transient regime where the laser energy is transferred to the plasma. 
Additionally, since the plasma response can lead to additional electromagnetic waves, results are shown from both a continuous laser and one with 6-periods only.  The latter allows us to observe re-emitted waves. 

Although we restrict this work to fluid-like resistivity and distribution functions, the Spitzer model generally needs to be modified for high- and low-frequency lasers \cite{dawson_highfrequency_1962}, as well as for varying ionization states \cite{skupsky_``coulomb_1987} and densities \cite{lee_electron_1984}. To maintain validity of the fluid model, the Langdon parameter \cite{langdon_nonlinear_1980, turnbull_impact_2020} here is $\alpha\approx1$, which keeps error from super-Gaussianity relatively low \cite{decker_nonlinear_1994}.

\begin{figure}[h!]
\centerline{\includegraphics[]{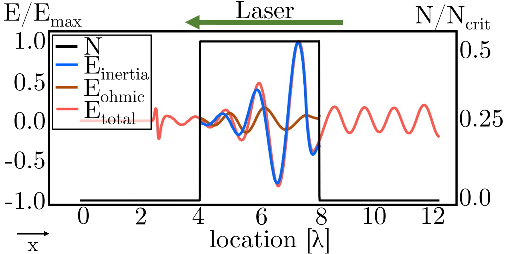}}
\caption{\label{fig:gol_vacuum_plasma} The electric field $\mathrm{E_z}$ output from the simulation compared to the components computed from Generalized Ohm's law.  Also plotted is the $\mathrm{N_{ion}}$ to show the validity across a plasma-vacuum interface. } 
\end{figure}

Fig.~\ref{fig:gol_vacuum_plasma} shows the electric field ($\mathrm{E_z}$) computed by the code in an initially isotropic under-dense step-density plasma.  This field is primarily composed of electron inertia ($\mathrm{\frac{\partial \mathbf{j}}{\partial t}}$) and the ohmic heating.  Initially, displacement currents dominate inside the low-density/vacuum portion of the plasma domain.  Then physical currents take over as the density is large enough to supply electrons easily.  Although this would not be a surprising result for a purely explicit 2-fluid code, we see that our semi-implicit 1-fluid (2-energy) XMHD formulation also captures this interaction.  

\begin{figure}[h!]
\centerline{\includegraphics[]{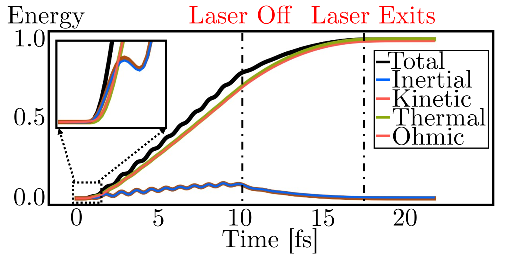}}
\caption{\label{fig:cumulative_energy_square_norm} The main cumulative contributions of the energy equation for electron fluids computed using $\mathrm{\mathbf{E}\cdot\mathbf{j}}$.  Also shown is the instantaneous partition of energy for electron fluid is as a sum of thermal ($\mathrm{k_bT}$) and directed kinetic ($\mathrm{mv^2}$).  The geometry here is a wedge-density profile pre-heated to 1 eV and with constant $\mathrm{ln\Lambda}=1$.  The laser power is $S=3.5\times10^{17}$ $\mathrm{W/m^2}$ ($a_0\approx0.003$).} 
\end{figure}

While there is collisional momentum transfer between ions and electrons, a detailed analysis found it relatively insignificant from the electron perspective. So, the dominant energy exchange in LPIs is best described by considering Poynting's theorem, 
\begin{equation}
   \mathrm{ \frac{d}{dt}(u_{E}+u_{B})+\nabla \cdot \mathbf{S}=-\mathbf{j}\cdot\mathbf{E}}.
   \label{eq:poynting_theorem}
\end{equation}
where $\mathbf{S}$ is the Poynting flux and $u_E$ and $u_B$ are the electric and magnetic energy densities, respectively.
The heating term on the RHS only gets turned into $\mathbf{E} \cdot \mathbf{j} \approx \eta j^2$ when $\mathbf{E}\approx \mathbf{\eta j}$.  However, the ohmic term in Eq. ~(\ref{eq:dimensionless_gol}) does not dominate the electric field in an under-dense plasma (see Fig.~\ref{fig:gol_vacuum_plasma}).  Instead, Fig.~\ref{fig:cumulative_energy_square_norm} shows the laser field heats the plasma by resonance of Eq. ~(\ref{eq:resonance_dimensionless}), forcing oscillatory kinetic energy into the electron fluid. This energy is then quickly thermalized through resistive collisions.  

The inset plot in Fig. \ref{fig:cumulative_energy_square_norm} shows the plasma first gains kinetic energy before resistivity initiates energy transfer. Under these assumptions the plasma behaves as an RLC circuit.  The oscillatory motion stores reactive energy, while the collisional energy corresponds to dissipation.  When the 6-period laser turns off, the remaining reactive oscillatory energy smoothly turns into thermal energy. $\mathrm{[\mathbf{E} \cdot \mathbf{j}]_{reactive}\approx \frac{m_e}{e^2N_e}j\frac{\partial j}{\partial t}=\frac{\partial{\frac{1}{2}\rho_e v_e^2}}{\partial t}}$ is just the kinetic energy given to the electron fluid by the laser.  If $\mathrm{ln\Lambda}$ is allowed to vary, then the plasma can store much more oscillatory energy as the increase in temperature lowers $\mathrm{ln\Lambda}$ and thereby reduces $\eta$ and increases the current density.   

\begin{figure}[!ht]
\centerline{\includegraphics[]{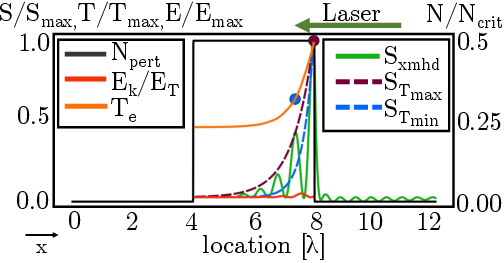}}
\caption{\label{fig:inverse_brehm_scaling}  The scaling factor between a basic inverse Bremsstrahlung model and the predictions from an XMHD model for two different laser powers with a plasma pre-heated to 1 eV (to ensure ionization), $\mathbf{S}=\mathrm{7.5\times10^{16}}$ $\mathrm{W/m^2}$. $\mathbf{S}$ refers to the Poynting flux, with the green curve computed from XMHD outputs, the dotted lines computed from analytic IB theory at two different temperatures, whose location in the plasma is shown in the filled-in circles.  The vertical axes are all unitless and refer to the fraction of maximum value for each quantity, except the density, which is the fraction of critical density.  Additionally, the red curve shows the fraction of directed kinetic energy ($\mathrm{\frac{E_k}{E_T}}$) in the system relative to the total energy (including thermal). } 
\end{figure}

We briefly show the model is also consistent with inverse Bremsstrahlung, where electromagnetic waves are damped at a rate $\mathrm{k_{ib} = \nu_{ei}\omega_{pe}^2/\mu c \omega^2}$, in the limit of of low laser intensity, $\mathbf{S}$.  Using a continuous laser, Fig.~\ref{fig:inverse_brehm_scaling} shows the Poynting flux, $\mathrm{\mathbf{S}_{xmhd}}$, computed directly from XMHD outputs along with the damped exponential function, $\mathrm{\mathbf{S}_{T_{min/max}}}$, computed from $\mathrm{k_{ib}}$.  However, a choice of what rapidly varying $T_e$ to use must be made when computing $\nu_{ei}$.  The color and spatial location of the solid circles show values and location used for corresponding curves, $\mathrm{\mathbf{S}_{T_{min}}}$ and $\mathrm{\mathbf{S}_{T_{max}}}$.  Although the match is nearly perfect for one choice of $T_e$, for higher laser intensity Fig.~\ref{fig:gol_vacuum_plasma} along with Eq. (\ref{eq:poynting_theorem}) show that since the electric field in the plasma is primarily composed of electron inertia for the first few wavelengths, then this leads to an increase in reactive kinetic energy over thermal kinetic energy.  This is a clear departure from IB theory.

\section*{Conclusions}
This letter explores how XMHD can be used to study the intricacies of collisional LPI across a broad range of parameters, that remains elusive to particle-in-cell and MHD models. It also demonstrates compatibility with both ray-tracing and particle-in-cell methods within their realms of validity.  While this work explores fundamental effects that are well understood, it does it by using a completely different approach. Leveraging the GOL, this work demonstrates XMHD can capture the oscillatory nature of the electrons and shows how the plasma frequency naturally arises in the proposed framework. We have shown that LPI appears as a mix of reactance, via resonance, and dissipation, via resistivity.  The reactive component stems from harmonic oscillations which stores mechanical energy inside the electron motion. This is only possible because electron have a non-zero mass in our model. The model can recover many fundamental aspects of LPI usually captured by particle-in-cell or multi-fluid MHD codes, such as reflection, absorption and even a critical density.  Energy transfer and its subsequent decay match well analytic theory for simple configurations, and at low energies.  Because of its accuracy, XMHD is appropriate for regimes that can be too expensive to simulate with multi-fluid codes and that MHD cannot capture.  

Since XMHD accounts for the entire $d\mathbf{j}/dt=\partial\mathbf{j}/\partial t+\mathbf{v}\cdot \nabla \mathbf{j}$, this framework naturally bridges the gap between the kinetic and hydrodynamics scales. It also ensures computational continuity between the short scales, dominated by electron physics, and the large scale, dominated by ion dynamics.  As a result, the effects of microscopic physics and their impact on microscopic scales can be done by using the same code. Under such assumptions, using one code to compute LPI and how they impact plasma dynamics on larger scales is clearly possible and will be explore in future work.

\section*{Funding}
This research was supported by the NSF CAREER Award PHY-1943939, the DOE center DE-NA0004148 and by the Laboratory for Laser Energetics Horton Fellowships.

\bibliography{bibliography}

\end{document}